\documentclass[preprint]{aastex}
\usepackage[]{epsfig}
\newcommand       \Angstrom     {\,{\rm \AA}}

\newcommand       \cm           {\,{\rm cm}}
\newcommand       \erg          {\,{\rm erg}}
\newcommand       \eV           {\,{\rm eV}}
\newcommand	  \g		{\,{\rm g}}
\newcommand	  \MV		{M_V}

\newcommand	  \Teff	        {T_{\rm eff}}
\newcommand	  \rmH		{{\rm H}}

\newcommand       \K            {\,{\rm K}}

\newcommand	  \pc		{\,{\rm pc}}
\newcommand	  \s		{\,{\rm s}}
\newcommand       \sr           {\,{\rm sr}}

\newcommand       \simlt        {\lesssim}

\newcommand	  \bc		{{b_{\rm C}}}
 
\newcommand       \vdb          {{\rm vdB\,133}}

\newcommand       \Urad         {U_{0}}

\newcommand	  \ppm		{\,{\rm ppm}}
\newcommand       \irratio      {[\lambda I_{\lambda}(12\micron)]_{\rm IRAS}/I_{\rm bol}(\rm FIR)}
\newcommand       \irtotot      {[\lambda I_{\lambda}(12\micron)]_{\rm IRAS}/I_{\rm bol}(\rm tot)}
\newcommand{\figwidth}{4.0in}

\countdef\decade=200
\advance\decade by \year
\countdef\hours=201
\advance\hours by \time
\divide\hours by 60
\countdef\mins=202
\advance\mins by \hours
\multiply\mins by 60
\multiply\hours by 100
\countdef\miltime=203
\advance\miltime by \hours
\advance\miltime by \time
\advance\miltime by -\mins

\shorttitle{On PAH Emission without UV Excitation}

\begin{document}

\title{
    \vspace*{-2.0em}
    {\normalsize\rm submitted to {\it The Astrophysical Journal Letters}}\\
    \vspace*{1.0em}
         Do the Infrared Emission Features Need Ultraviolet Excitation?\\
         The PAH Model in UV-poor Reflection Nebulae\\
	 }

\author{Aigen Li and B.T. Draine}
\affil{Department of Astrophysical Sciences, Princeton University, 
       Peyton Hall, Princeton, NJ 08544, USA;\\
        {\sf agli@astro.princeton.edu, draine@astro.princeton.edu}}

\begin{abstract}
One of the major challenges to identification of the 3.3, 6.2, 7.7, 
8.6, and 11.3$\micron$ interstellar infrared (IR) emission bands 
with polycyclic aromatic hydrocarbon (PAH) molecules has been 
the recent detection of these bands in regions with little 
ultraviolet (UV) illumination, since small, neutral PAH molecules 
have little or no absorption at visible wavelengths 
and therefore require UV photons for excitation.

We show here that our ``astronomical'' PAH model, 
incorporating the experimental result that 
the visual absorption edge shifts to 
longer wavelength upon ionization and/or 
as the PAH size increases, can closely reproduce 
the observed infrared emission bands 
of vdB 133, a UV-poor reflection nebula.

It is also shown that single-photon heating of ``astronomical'' PAHs in
reflection nebulae near stars as cool as $\Teff=3000\K$
can result in observable emission
at 6.2, 7.7, 8.6, and 11.3$\micron$.
Illustrative mid-IR emission spectra are also calculated
for reflection nebulae illuminated by
cool stars with $\Teff=3500, 4500, 5000\K$.
These will allow comparison with future 
Space Infrared Telescope Facility (SIRTF) observations 
of vdB 135 ($\Teff=3600\K$), vdB 47 ($\Teff=4500\K$), 
and vdB 101 ($\Teff=5000\K$).

It is also shown that the dependence of the 12$\micron$ IRAS emission 
relative to the total far-IR emission on the effective temperature of 
the exciting star is consistent with the PAH model expectation
for $3000\K \le \Teff \le 30000\K$.  

\end{abstract}


\section{Introduction\label{sec:intro}}
Since their first detection in the planetary nebulae NGC 7027 and 
BD+30$^{\rm o}$3639 (Gillett, Forrest, \& Merrill 1973), 
the distinctive set of infrared (IR) emission features at
3.3, 6.2, 7.7, 8.6, and 11.3$\micron$ have been observed in 
a variety of objects with a wide range of physical conditions, 
including planetary nebulae, protoplanetary nebulae, 
reflection nebulae, HII and ultracompact HII regions, 
circumstellar envelopes, the diffuse interstellar medium (ISM)
of the Milky Way Galaxy, and external galaxies 
(see Tielens et al.\ 1999 and Sellgren 2001 for recent reviews).
Although the exact nature of their carriers remains unidentified
-- they remain known as ``the Unidentified
Infrared (UIR) bands'' -- it is now widely thought that they
originate from some sorts of aromatic hydrocarbons. 
Various carbonaceous materials have been proposed as 
the UIR band carriers. In general, they can be categorized 
into two classes: (1) pure, free-flying aromatic molecules -- 
polycyclic aromatic hydrocarbon molecules (PAHs; L\'{e}ger \& Puget 1984; 
Allamandola, Tielens, \& Barker 1985);
(2) carbonaceous grains with a partly aromatic character 
-- hydrogenated amorphous carbon (HAC; Duley \& Williams 1981; 
Jones, Duley, \& Williams 1990), 
quenched carbonaceous composite (QCC; Sakata et al.\ 1990), 
coal (Papoular et al.\ 1993), fullerenes (Webster 1993), 
and surface-graphitized nanodiamonds (Jones \& d'Hendecourt 2000).

In most current models, the UIR band emission involves 
three sequential steps: 
(1) excitation by absorption of an energetic starlight photon
(usually ultraviolet [UV]);
(2) rapid ($\sim 10^{-12}-10^{-10}\s$) redistribution of all or part
of the absorbed photon energy 
over all available vibrational modes;
(3) radiative relaxation via IR fluorescence. 

Among the existing proposed carriers, the PAH model is gaining
increasing acceptance because of 
(1) the close resemblance of the UIR spectra (frequencies and 
relative intensities) to the vibrational spectra of PAH molecules 
(e.g. see Allamandola, Hudgins, \& Sandford 1999) and 
(2) the ability of a PAH molecule to emit efficiently 
in the UIR wavelength range following single photon heating 
(Greenberg 1968; L\'{e}ger \& Puget 1984; 
Allamandola, Tielens, \& Barker 1985; Draine \& Li 2001).
In contrast, larger carbonaceous grains are unlikely to be 
efficient UIR emitters since the timescale for the absorbed 
photon energy to diffuse ($\sim 10^{-9}\s$)
is much shorter than the IR vibrational emission 
timescale ($\sim 0.1\s$; see Tielens et al.\ 1999). 

Early observations of the UIR emission bands were made in regions 
with strong UV irradiation, and PAH excitation was 
expected since
all PAH species are strongly absorbing 
in the vacuum ultraviolet ($\lambda \simlt 3000$\AA).
The PAH UV excitation model has recently been challenged
by the ISO ({\it Infrared Space Observatory}) detection 
of the UIR bands in $\vdb$, a reflection nebula illuminated 
by a binary system with little UV radiation 
(see Uchida, Sellgren, \& Werner 1998).
The UIR spectrum of this UV-poor region closely resembles
those observed in sources with much harsher UV environments 
(Uchida et al.\ 2000). 
This appears to be in conflict with the view that 
PAHs are primarily excited by UV photons,
as would be expected based on laboratory studies 
showing that the absorption by {\it small, neutral} PAHs 
has a sharp cut-off 
in the UV, with little or no absorption in the visible
(see Sellgren 2001 for a review).  
Uchida et al.\ (2000) obtained 5--15$\micron$ ISOCAM spectra
of 6 reflection nebulae and found no systematic spectroscopic 
differences despite values of $\Teff$ ranging from 6800$\K$ 
to 19000$\K$.

The PAH electronic absorption edge is known to shift to 
longer wavelength with increasing size and/or 
upon ionization (see Allamandola et al.\ 1989, Salama et al.\ 1996 
and references therein).
While the largest experimentally studied PAH molecule
to date is dicoronylene C$_{48}$H$_{20}$ 
(Allamandola, Hudgins, \& Sandford 1999),
astronomical PAHs are believed to be larger 
(e.g., the mean size for the Milky Way PAHs 
is $\approx 6$\AA, corresponding to $N_{\rm C}\approx 100$
[Li \& Draine 2001a]). 
The astronomical PAH model --
with the size/ionization dependence of the PAH absorption edge 
taken into account (see \S A2 in Li \& Draine 2001a) --
is successful in explaining the observed mid-IR spectra 
of the Milky Way diffuse ISM (Li \& Draine 2001a), 
the quiescent molecular cloud SMC B1\#1 
in the Small Magellanic Cloud (Li \& Draine 2001b),
and the UIR band ratios for a wide range of environments ranging 
from reflection nebulae, HII regions, photodissociation regions (PDRs),
molecular clouds in the Milky Galaxy to normal galaxies,
starburst galaxies, and a Seyfert 2 galaxy (Draine \& Li 2001). 

In this {\it Letter} we show that the astronomical PAH model is 
consistent with the observed UIR emission from UV-poor environments. 
In \S\ref{sec:vdb133} we verify that the astronomical PAH model
can quantitatively reproduce the vdB 133 UIR spectrum.
We further demonstrate in \S\ref{sec:coolrn} that
the UIR bands are also expected for reflection nebulae 
which are even more UV-poor than vdB 133,
and we provide model spectra for comparison with 
future {\it Space Infrared Telescope Facility} 
(SIRTF) observations of reflection nebulae near cool stars.
In \S\ref{sec:12um} we show that the predicted $\Teff$ dependence 
of the ratio of the IRAS 12$\micron$ emission 
to the total far-IR emission is also consistent with observations.
Our results are discussed in \S\ref{sec:discuss},
and our conclusions are summarized in \S\ref{sec:sum}.

\section{vdB 133 \label{sec:vdb133}}

The reflection nebula vdB 133, located in the Cyg OB1 association 
at a distance $d\approx 1500\pc$, is illuminated by 
a binary system consisting of HD 195593A (spectral type F5Iab;
effective temperature $\Teff \approx 6800\K$;
absolute magnitude $\MV=-6.6\pm 0.4$)
and HD 195593B (spectral type B7II; $\Teff \approx 12000\K$;
$\MV=-4.0\pm 0.4$)
(see Uchida et al.\ 2000 and references therein). 
Following Uchida et al.\ (1998), we approximate the vdB 133 
radiation field by a linear combination of two Kurucz model 
atmosphere spectra (Kurucz 1979).
The starlight intensity incident on the reflection nebula is assumed to
have
$U_0\approx 135$, where $U_0$ is the
ratio of the 912\AA--1$\micron$ energy density relative to the
value for the MMP interstellar radiation field 
($4.80\times10^{-13}\erg\cm^{-3}$).

Let $f_\lambda$ 
be the fraction of the nonionizing stellar luminosity $L_\lambda$
radiated at wavelength $< \lambda$:
\begin{equation}
f_\lambda \equiv \frac{\int_{912\Angstrom}^{\lambda} L_{\lambda^\prime} 
d\lambda^\prime}
{\int_{912\Angstrom}^{\infty}L_{\lambda^\prime} d\lambda^\prime}
\end{equation}
This binary system,\footnote{%
	HD 195593B, with $f_{\rm 0.2\mu m} \approx 32\%$,
	contributes $\approx 80\%$ of the total flux shortward of 0.2$\micron$;
	HD 195593A, with $f_{\rm 0.2\mu m} \approx 0.4\%$,
	contributes $\approx 95\%$ of the total flux shortward of 1$\micron$.
	}
with $f_{\rm 0.2\mu m} \approx 1.7\%$,
has the lowest value of $f_{0.2\mu{\rm m}}$
for reflection nebulae with detected UIR emission bands.\footnote{
	For comparison, NGC 2023 is illuminated by HD 37903, with
	$\Teff=22000\K$ and $f_{\rm 0.2\mu m} \approx 68\%$;
	the average interstellar starlight background 
	(Mathis, Mezger, \& Panagia 1983) has 
	$f_{\rm 0.2\mu m}=4.7\%$.
	}
Despite this, the vdB 133 UIR features
are very similar to those observed in regions with much harder
radiation fields (Uchida et al.\ 2000).

\begin{figure}[h]
\begin{center}
\epsfig{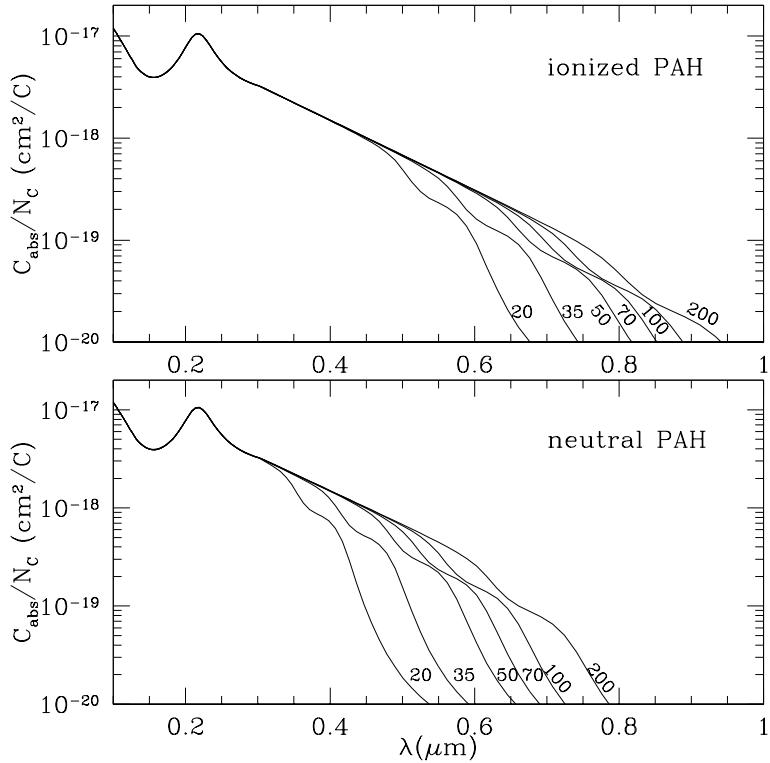}
\end{center}\vspace*{-1em}
\caption{
        \label{fig:pahcabs}
        \footnotesize
        Absorption cross sections (per C atom)
	estimated for 
	astronomical PAH molecules containing
        $N_{\rm C}=20,35, 50, 70, 100, 200$ C atoms, for
        both charged (upper panel) 
        and neutral (lower panel) states.
        }
\end{figure}

We model the vdB 133 UIR features as emitted from a mixture of
neutral and ionized PAH molecules. 
We use the PAH absorption cross sections estimated
by Li \& Draine (2001a) which take into account
the size/charge dependence of the PAH long wavelength 
absorption edge (see Figure \ref{fig:pahcabs}).
We adopt a log-normal size distribution for the PAHs, 
characterized by three parameters: $a_{0}$, $\sigma$, and $\bc$;
$a_{0}$ and $\sigma$ respectively determine 
the peak location and the width of the log-normal distribution,
and $\bc$ is the total amount of C atoms 
(relative to total H) locked up in PAHs. 
As always, the term ``PAH size'' refers to the radius $a$ 
of a hypothetical spherical grain with the same carbon density 
as graphite ($2.24\g\cm^{-3}$) and containing
the same number of carbon atoms $N_{\rm C}$: 
$a \equiv 1.288 N_{\rm C}^{1/3}\Angstrom$. 

The PAH ionization fraction depends on
the radiation intensity, the electron density $n_e$,
and the gas temperature $T_{\rm gas}$. 
Assuming
$T_{\rm gas}=100\K$,
we estimate the PAH ionization fractions in vdB 133 
using the rates for photoelectric emission and 
electron capture recently discussed by Weingartner \& Draine (2001).

We employ the ``thermal-discrete'' method (Draine \& Li 2001)
to calculate the temperature distribution functions
for PAHs subject to single-photon heating.
Our model assumes a single radiation field: 
attenuation of the illuminating
starlight in the dust layer is neglected.
If the dust layer is optically-thick to the exciting radiation,
we may approximate this by an optically-thin layer with a column
density $N_\rmH$ such that the optical depth to the exciting
radiation would be unity.  For a standard dust mixture and
$\lambda = 4000\Angstrom$ (say), this would correspond to
$N_\rmH\approx 1.3\times10^{21}\cm^{-2}$.

\begin{figure}[h]
\begin{center}
\epsfig{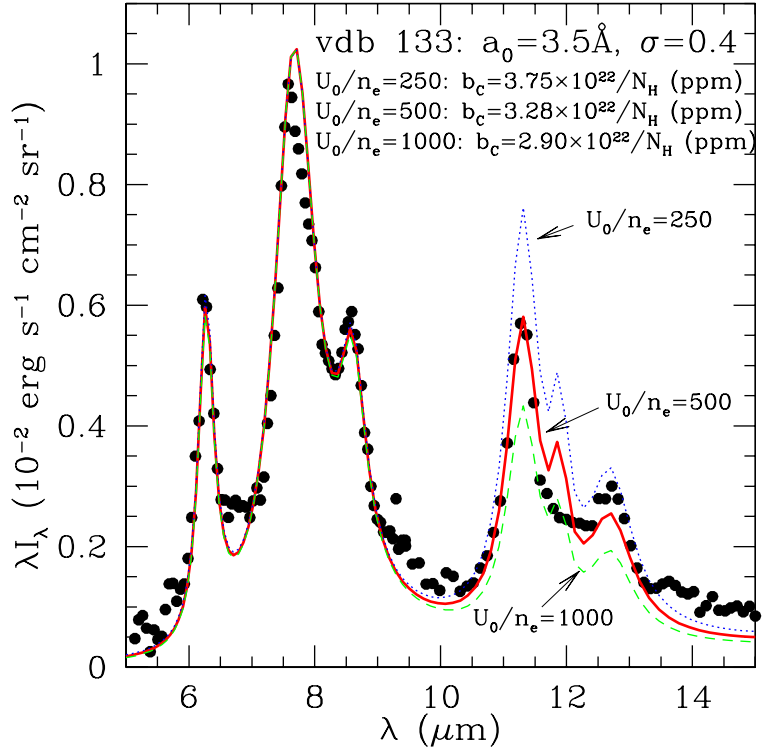}
\end{center}\vspace*{-1em}
\caption{
        \label{fig:vdb133}
        \footnotesize
        Predicted IR emission spectra for PAHs in vdB 133 
        in comparison with the ISOCAM observational spectrum 
        (filled circles; Uchida et al.\ 1998).
        The PAH mixture is modeled as a log-normal
        distribution of $a_0=3.5$\AA, $\sigma=0.4$,
        $\bc \approx 38\ppm\times (10^{21}\cm^{-2}/N_{\rm H})$
        for $\Urad/n_e=250\cm^3$ (dotted line),
        $\bc \approx 33\ppm\times (10^{21}\cm^{-2}/N_{\rm H})$
        for $\Urad/n_e=500\cm^3$ (solid line),
        and $\bc \approx 29\ppm\times (10^{21}\cm^{-2}/N_{\rm H})$
        for $\Urad/n_e=1000\cm^3$ (dashed line).
        We take the observed UIR band widths 
        (see Table 7 of Li \& Draine 2001).
        }
\end{figure}

Figure \ref{fig:vdb133} plots the theoretical spectra 
calculated from models with $a_0=3.5$\AA, $\sigma=0.4$, and
abundances $(N_\rmH/10^{21}\cm^{-2})\bc = 38, 33, 29\ppm$
for $\Urad/n_e=250, 500, 1000\cm^3$ respectively.
We see that the $\Urad/n_e=500\cm^3$ model ($n_e=0.3\cm^{-3}$)
closely reproduces
the observed UIR spectrum using the standard Milky Way PAH size 
distribution ($a_0=3.5$\AA, $\sigma=0.4$),
a realistic radiation field,
and PAH abundances ($\bc\approx 33\ppm$ if $N_\rmH=10^{21}\cm^{-2}$)
comparable to the Milky Way abundances ($\bc\approx 45\ppm$).

\section{Cool Reflection Nebulae \label{sec:coolrn}}

We have seen in \S\ref{sec:vdb133} that the astronomical PAH
model is successful in explaining the UIR spectrum observed
in vdB 133. 
In this section we will show that reflection nebulae illuminated by
{\it cooler} stars are also capable of exciting PAHs sufficiently
to emit at the 6.2, 7.7, 8.6 and 11.3$\micron$ UIR bands.
For illustrative purpose, we consider reflection nebulae 
illuminated by stars with $\Teff = 3000, 3500, 4500, 5000\K$. 
The latter 3 cases will serve as a direct comparison basis for
SIRTF observations of vdB 135 ($\Teff=3600\K$),
vdB 47 ($\Teff=4500\K$),
and vdB 101 ($\Teff=5000\K$) (Houck 2001).\footnote{%
 Attempts by Uchida et al.\ (2000) to search for UIR bands in
 vdB 101 and vdB 135 using ISOCAM spectroscopy
 were unsuccessful due to the faintness of these nebulae.
 }

\begin{figure}[h]
\begin{center}
\epsfig{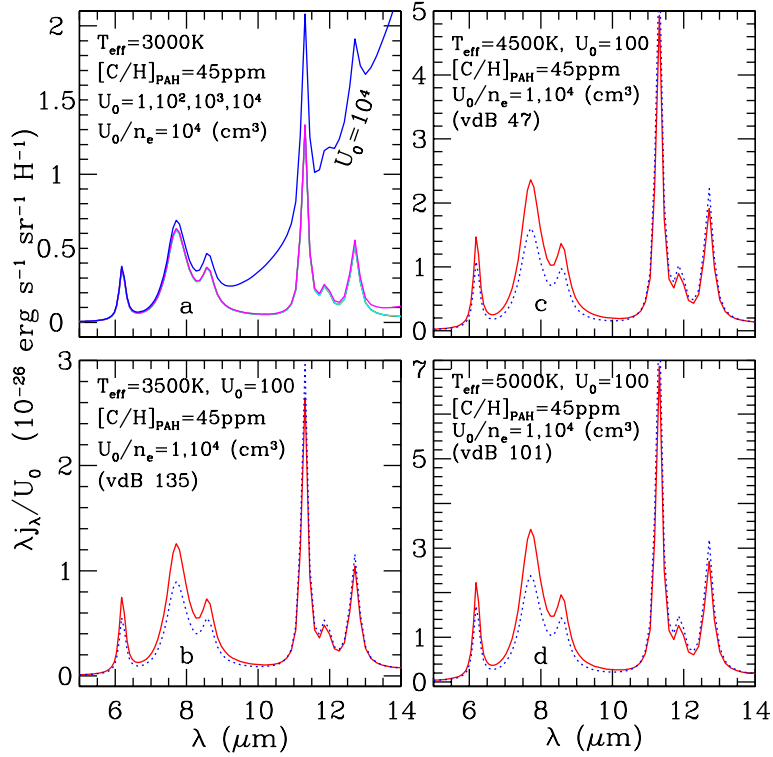}
\end{center}\vspace*{-1em}
\caption{
        \label{fig:pahcoolrn}
        \footnotesize
        Predicted 5--14$\micron$ emission spectra for
        reflection nebulae illuminated by cool stars with
        an effective temperature of $\Teff = 3000\K$ and 
        starlight intensity $\Urad=1, 100, 1000, 10^4$ (a), 
        $\Teff=3500\K$ and $\Urad=100$ (b), 
        $\Teff=4500\K$ and $\Urad=100$ (c), 
    and $\Teff=5000\K$ and $\Urad=100$ (d). 
        In all models the PAH size distributions are taken to
        be that of the Milky Way diffuse ISM: 
        a log-normal size distribution with $a_0=3.5$\AA,
        $\sigma=0.4$, and $\bc=45\ppm$.
        We take $n_e/n_{\rm H}=1.2\times 10^{-4}$, $T_{\rm gas}=100\K$,
        and $\Urad/n_e=1.0\cm^3$ (solid lines)
        or $\Urad/n_e=10^4\cm^3$ (dotted lines)
        for calculating the PAH charging. 
        }
\end{figure}

Lacking prior knowledge of the PAH size distribution
and the physical environmental parameters such as $n_e$,
$T_{\rm gas}$, and the starlight intensity $\Urad$,
we adopt the Milky Way PAH size distribution
(i.e., a log-normal distribution with $a_0=3.5$\AA\ and $\sigma =0.4$
[Li \& Draine 2001])
and we take $T_{\rm gas}=100\K$.
We take $\Urad=100$ as a representative number since 
$30\simlt \Urad \simlt 300$ for typical reflection nebulae   
(see Sellgren et al.\ 1990);
the calculated UIR emission is essentially proportional to $\Urad$
for $\Urad \simlt 10^4$ 
(see Figure \ref{fig:pahcoolrn}a). In all models we take the
number of carbon atoms relative to hydrogen in PAHs to be
$\bc=45\ppm$, the same as in the Milky Way diffuse ISM (Li \& Draine 2001).

For stars with $\Teff\ge 3500\K$ we take the Kurucz atmospheric
model spectra to approximate the radiation fields. 
Since the Kurucz model is not available for stars cooler than 3500$\K$,
we take the model spectrum of Allard et al.\ (1997) for $\Teff=3000\K$.

In Figure \ref{fig:pahcoolrn}a we plot the mid-IR spectra calculated
for PAHs heated by a $\Teff=3000\K$ star with $\Urad=1, 100, 10^3, 10^4$
and $\Urad/n_e=10^4$.
The predicted PAH IR bands are prominent, 
with a much stronger 11.3$\micron$ C-H out-of-plane band 
in comparison with those in regions illuminated by much hotter stars 
(e.g., NGC 2023 [$\Teff=22000\K$], Orion Bar [$\Teff=37000\K$], 
M17 [$\Teff=45000\K$]; Verstraete et al. 2001).
This is due to the lack of UV photons capable of 
causing photoelectric ejection of electrons and 
thus PAHs in such a cool region mainly remain neutral
even for $\Urad/n_e=10^4$.
As expected from the single-photon heating scenario,
the PAH IR features are almost indistinguishable for $\Urad\simlt 10^4$
except the $\Urad=10^4$ model has a rising wing for $\lambda > 10\micron$. 
This is because for $\Urad=10^4$ the $>20\Angstrom$ dust grains
already attain an equilibrium temperature $T\approx 100\K$.

In Figure \ref{fig:pahcoolrn}b,c,d we plot the model spectra
for $\Urad=100$, $n_e=0.01, 100\cm^{-3}$,
and $\Teff=3500, 4500, 5000\K$, respectively. 
It is seen that all spectral shapes are almost the same except 
the absolute emissivity level (at fixed $\Urad$) 
increases with increasing of $\Teff$. 
This is because hotter stars radiate a larger fraction of their power
at short wavelengths where PAHs are strongly absorbing.
It is also seen from the enhanced 11.3$\micron$ band relative to
the 7.7$\micron$ C-C stretching band that a large fraction of PAHs 
in those regions are neutral.\footnote{%
 The probability for a small PAH molecule being neutral 
 is larger for $\Urad/n_e=10^4\cm^3$ than that for $\Urad/n_e=1.0\cm^3$.
 This is because in regions devoid of UV photons 
 the charging of a neutral PAH is dominated by electron capture
 and thus PAHs are more likely to be {\it negatively} charged when
 $\Urad/n_e$ is small.
 }
This is the major difference between the UIR features excited
by cool stars with those excited by hot stars.\footnote{%
 The vdB 133 nebula is not lack of photoelectron-ejecting photons. 
 The hot companion HD 195593B ($\Teff=12000\K$)
 of the exciting binary system contributes $\approx 80\%$ 
 of the total 6--13.6$\eV$ photons.
 This is why the vdB 133 UIR spectrum resembles 
 those excited by hot stars.
 }
It is hoped that SIRTF observations of these reflection nebulae
will provide both spectra of the UIR bands and a full
spectral energy distributions to allow the illuminating intensity
$U_0$ to be well-determined, thereby allowing comparison with the model
spectra in Figure \ref{fig:pahcoolrn}.

In Table \ref{tab:coolrn} we give the PAH mid-IR emission
convolved with the 5.8$\micron$ and 8.0$\micron$ SIRTF 
{\it Infrared Array Camera} (IRAC) bands
calculated for the Milky Way PAH mixture excited by cool stars
with $\Teff=3000, 3500, 4500, 5000\K$. Figure \ref{fig:pahcoolrn}
and Table \ref{tab:coolrn} may provide guidance for the coming
SIRTF observation of cool reflection nebulae (Houck 2001). 
We stress, however, that the actual mid-IR spectra 
may differ from those in Figure \ref{fig:pahcoolrn} and Table \ref{tab:coolrn}
since the PAH size distribution, the PAH abundance,
and the environmental parameters $n_e$, $T_{\rm gas}$ and
$\Urad$ may differ from those adopted here.

\begin{table}[h]
\begin{center}
\caption[]{PAH emissivity averaged over the 5.8$\micron$ 
           and 8.0$\micron$ SIRTF IRAC bands for 
           the Milky Way mixture
           illuminated by cool stars
           for $\Urad/n_e=1.0\ (10^4)\cm^3$\label{tab:coolrn}.}
\begin{tabular}{ccccc}
\hline
\hline
$\lambda_{\rm eff}$ & \multicolumn{4}{c}{$\frac{1}{\Urad}
\left(\frac{\lambda I_{\lambda}}{N_{\rm H}}\right)
\left(\frac{45\ppm}{\bc}\right)$ 
(${\rm 10^{-27}\ erg\ s^{-1}\ \sr^{-1}\ H^{-1}}$)}\\
\cline{2-5}
($\micron$) & $\Teff=3000\K$ & $\Teff=3500\K$ & $\Teff=4500\K$ 
                    & $\Teff=5000\K$ \\
\hline
5.8	        & 0.64 (0.49)
		& 1.34 (1.02)
		& 2.66 (2.01)
		& 4.05 (3.17)\\
8.0  	        & 2.69 (2.04)
	        & 5.41 (3.93)
	        & 10.1 (7.05)
	        & 14.7 (10.5)\\
\hline
\end{tabular}
\end{center}
\end{table}

\section{The 12$\micron$ Emission vs. Radiation Hardness\label{sec:12um}}

To study the carrier of the 12$\micron$ emission and its excitation
mechanism, Sellgren et al.\ (1990) investigated how
the ratio 
$[\lambda I_{\lambda}(12\micron)]_{\rm IRAS}/I_{\rm bol}(\rm FIR)$
depended
on the effective temperature $\Teff$
of the exciting stars for 24 reflection nebulae,
where $[\lambda I_{\lambda}(12\micron)]_{\rm IRAS}$ is estimated from
the IRAS 12$\micron$ filter, and 
the total far-IR surface brightness
$I_{\rm bol}(\rm FIR)$ was estimated by fitting
a blackbody multiplied
by a $1/\lambda$ emissivity to the IRAS 60 and 100$\micron$
surface brightnesses.
They found that $\lambda I_{\lambda}(12\micron)/I_{\rm bol}(\rm FIR)$
is basically independent of $\Teff$ for $5000\K\le \Teff \le 33000\K$.
If PAHs are the 12$\micron$ carrier and excited only by UV photons, 
one would expect $\irratio$ to show a sharp drop for cool stars
($\Teff < 10000\K$) which are relatively devoid of UV radiation  
(Sellgren et al.\ 1990). Therefore, one can conclude that either
PAHs are not responsible for the 12$\micron$ emission or PAHs
are excited by photons of a wide range of wavelengths.   
Since it has been well established that the UIR features 
are the dominant contributor to the 12$\micron$ emission,
the former hypothesis can be ruled out. 
Therefore, it must be possible for photons longward of the UV to excite
the UIR carrier.  Is this consistent with our understanding of PAHs?

\begin{figure}[h]
\begin{center}
\epsfig{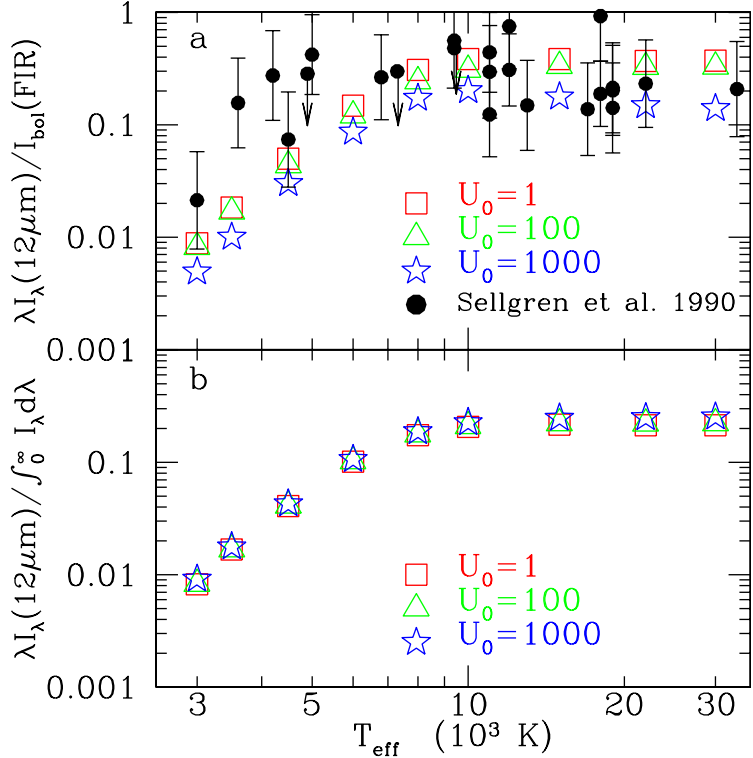}
\end{center}\vspace*{-1em}
\caption{
        \label{fig:rn12um2fir}
        \footnotesize
        Ratio of the 12$\micron$ emission 
        to the total far-IR emission estimated from the $60$ and $100\micron$
	IRAS fluxes (upper panel; see text)
        and to the actual total absorbed energy (lower panel).
        Observational data (filled circles) are taken from
        Sellgren, Luan, \& Werner (1990).
        Model results are for reflection nebulae illuminated by
        stars with $T_{\rm eff} = 3000, 3500, 4500, 6000, 8000,
        10000, 15000, 22000, 30000\K$ and starlight intensity
        $\Urad=1, 100, 1000$ times that of 
        the 912\AA--1$\micron$ MMP interstellar radiation field,
        assuming $\Urad/n_e=100\cm^3$ and    
        using the Milky Way dust mixture.
        }
\end{figure}

To test the astronomical PAH model, we calculate $\irratio$ 
for reflection nebulae illuminated by stars with 
$\Teff$=3000, 3500, 4500, 6000, 8000, 10000, 15000, 22000
and 30000$\K$ and starlight intensity $\Urad=1, 100, 1000$,
assuming $\Urad/n_e=100\cm^3$ and using the Milky Way dust mixture.
As shown in Figure \ref{fig:rn12um2fir}a the model results
are in good agreement with observational data except 
four nebulae: vdB 135 ($\Teff=3600\K$), vdB 42 ($\Teff=4200\K$),  
vdB 47 ($\Teff=4500\K$), vdB 35 ($\Teff=4900\K$).
It is worth noting that these nebulae are the four faintest 
among Sellgren et al.'s sample of 24. It is possible 
that $I_{\rm bol}(\rm FIR)$ was underestimated for these
cool, faint nebulae as a consequence of the method 
of obtaining $I_{\rm bol}(\rm FIR)$ because 
the 60 and 100$\micron$ emission preferentially account the warm
grain component while the bulk of the far-IR emission is probably
longward of 100$\micron$ for cool reflection nebulae.

A better way to characterize the fraction of total IR emission
emitted at 12$\micron$ is to replace $I_{\rm bol}(\rm FIR)$
by $I_{\rm bol}(\rm tot)\equiv \int_{0}^{\infty} I_{\lambda} d\lambda$.
In Figure \ref{fig:rn12um2fir}b we show the model predicted
$\irtotot$ as a function of $\Teff$. 

\section{Discussion\label{sec:discuss}}

It is not surprising that the astronomical PAH model is
able to reproduce the vdB 133 UIR spectrum. It is also
not unexpected that the PAH mid-IR emission spectra in regions
devoid of UV radiation resemble those in UV-rich regions.
This can be readily seen from Figure 7 of Draine \& Li (2001) 
which shows that a 0.5$\micron$ photon will
excite a $N_{\rm C}=100$ (mean size for the Milky Way 
mixture) PAH molecule to a vibrational temperature of 
$T\approx 390\K$, resulting in strong emission in
the 6.2--11.3$\micron$ UIR bands. 
For smaller PAHs, softer photons
are able to fulfill this excitation task.
Therefore, the detection of UIR bands in UV-poor regions 
does not contradict PAHs as the UIR band carrier.
The lack of a precipitous drop for cool stars
($\Teff < 10000\K$) in $\irratio$ simply indicates that
interstellar PAHs have absorption extending into the visible,
presumably as the combined result of somewhat larger sizes than 
those currently studied in laboratory, plus modification of 
the absorption properties by ionization.

Papoular (2000) proposed that PAH excitation could result from
energy released during PAH-catalyzed H$_2$ formation.
While this excitation mechanism may be present,
it does not appear able to provide sufficient excitation to
account for the observed UIR emission (Pagani et al.\ 1999).
In any case, our model calculations show that
excitation by starlight can account for the observed
UIR intensities.

We note that our approach to estimating the absorption properties of
astronomical PAHs was simplified (Li \& Draine 2001) in 
that we did not distinguish negatively-charged anions (PAH$^-$) 
and multiply-charged cations (PAH$^{n+}$) 
and anions (PAH$^{n-}$) from singly-charged cations (PAH$^+$). 
For example,
laboratory studies of the electronic absorption spectra
of coronene (C$_{24}$H$_{12}$) in solution have shown that 
its di-anion (C$_{24}$H$_{12}^{2-}$) has markedly enhanced
visual/near-IR continuum absorption than its 
mono-anion (C$_{24}$H$_{12}^{-}$) and neutral counterpart
(Hoijtink 1959). 
However, the fact that our simplified PAH model can successfully
reproduce the vdB 133 UIR spectrum and the $\Teff$
dependence of $\irratio$ suggests that the model provides
a fairly good approximation to the real absorption properties
of astronomical PAH mixtures.

\section{Conclusion\label{sec:sum}}

We have modeled the excitation of PAH molecules in UV-poor
regions. It is shown that the astronomical PAH model provides
a satisfactory fit to the UIR spectrum of vdB 133, 
a reflection nebulae with the lowest ratio of UV to total
radiation among reflection nebulae with UIR bands detected. 
It is also shown that astronomical PAHs can be pumped
by cool stars with even less UV radiation. 
It is further shown that
the PAH model predicts a dependence
of $\irratio$ for reflection nebulae which 
is consistent with observations
for $3000\K \le \Teff \le 30000\K$.  
We conclude that PAHs appear able to
account for the UIR band emission
observed in reflection nebulae.

\acknowledgments
We thank K.I. Uchida for providing us with the ISO spectrum 
of vdB 133, F. Allard, C. Dominik and P.H. Hauschildt for their
help in obtaining stellar model spectra, 
and R.H. Lupton for the availability of the SM plotting package. 
We thank L.J. Allamandola, J.M. Greenberg, K. Sellgren 
and K.I. Uchida for helpful discussions.
This research was supported in part by 
NASA grant NAG5-10811 and NSF grant AST-9988126.

\end{document}